\documentclass[sn-mathphys-num]{sn-jnl}% Math and Physical Sciences Numbered Reference Style 
\usepackage{graphicx}%
\usepackage{multirow}%
\usepackage{amsmath,amssymb,amsfonts}%
\usepackage{amsthm}%
\usepackage{mathrsfs}%
\usepackage[title]{appendix}%
\usepackage{xcolor}%
\usepackage{textcomp}%
\usepackage{manyfoot}%
\usepackage{booktabs}%
\usepackage{algorithm}%
\usepackage{algorithmicx}%
\usepackage{algpseudocode}%
\usepackage{listings}%

\theoremstyle{thmstyleone}%
\theoremstyle{thmstyletwo}%

\theoremstyle{thmstylethree}%

\begin{document}

\title[Mendeleev Periodic Table]{Research status of the Mendeleev Periodic Table: a bibliometric analysis}

\author[1]{\fnm{Kamna} \sur{ Sharma}}\email{kamna.sharma\_phd23@gla.ac.in}

\author[1]{\fnm{Deepak Kumar} \sur{Das}}\email{deepak.das@gla.ac.in}
\equalcont{These authors contributed equally to this work.}

\author*[2]{\fnm{Saibal} \sur{Ray}\footnote{Orcid ID: 0000-0002-5909-0544}}\email{saibal.ray@gla.ac.in}
\equalcont{These authors contributed equally to this work.}

\affil[1]{\orgdiv{Department of Chemistry}, \orgname{GLA University}, \orgaddress{\street{Street}, \city{Mathura}, \postcode{281406}, \state{Uttar Pradesh}, \country{India}}}

%\affil[1]{\orgdiv{Department of Chemistry}, \orgname{GLA University}, \orgaddress{\street{Street}, \city{Mathura}, \postcode{281406}, \state{Uttar Pradesh}, \country{India}}}

\affil[2]{\orgdiv{Centre for Cosmology, Astrophysics and Space Science (CCASS)},  \orgname{GLA University}, \orgaddress{\street{Street}, \city{Mathura}, \postcode{281406}, \state{Uttar Pradesh}, \country{India}}}

\abstract{In this paper, we present a bibliometric analysis of the Mendeleev Periodic Table. We have conducted a comprehensive analysis of Scopus based database using the keyword ``Mendeleev Periodic Table". Our findings suggest that the Mendeleev Periodic Table is an influential topic in the field of Inorganic as well as Organic Chemistry. Future researcher may focus on expanding our analysis to include other bibliometric indicators to gain a more comprehensive understanding of the impact of the Mendeleev Periodic Table in the chemistry-based scientific investigations and even in the field of astrochemistry.}


\maketitle

\section{Introduction}\label{sec1}

~~~Bibliometrics is the use of statistical methods to analyse books, articles and other publications, especially in scientific contents. Bibliometric methods are frequently used in the field of library and information science as a tool for finding the impact of publications using statistics. It involves collecting data, processing it and analysing from bibliographic database like Scopus, Web of Science or Google Scholar. In a way we can say that bibliometric analysis can be used to explore and analyse large volume of scientific data, unpack the evolutionary nuances of specific field and shed light on emerging areas in that field. There are several tools available for conducting analysis on bibliometric. It is an R tool for comprehensive science mapping analysis. 

Therefore, other than library science~\cite{CRC2009,Coombs2017,Cabezas2021} and business as well as administrative management~\cite{Evans1993,Schaufeli1996,Maslach1998,Leiter2001,Schaufeli2002,Sharma2002,Kumar2013,Wang2019,Bakera2020} nowadays there is a growing trend to exploit this effective tool in the other field also, e.g. ecology and evolutionary biology~\cite{Brown2018,Claireaux2018,Pesendorfer2019}, psychology~\cite{Figueredo2006,Griskevicius2011,Sacco2012,Del2014}, anthropology~\cite{Kaplan2000}, public health~\cite{Wells2017}, criminology~\cite{Dunkel2013}, accountancy~\cite{Leiby2017}, economics~\cite{Zeng2023} and so on. Recently a successful application of bibliometrics has been noted in the biological system~\cite{Nettle2019} where the authors used a bibliometric analysis of an interdisciplinary research area in the context of the evolution of life-history theory.  

Motivated by the above mentioned works based on the bibliometric analysis we have performed a study in the purview of basic science, especially in the field of chemistry. However, at this juncture we would like to recall the comment by the famous astrophysicist Carl Sagan who relates the basic chemistry to the astrophysics in the following unique manner: “The nitrogen in our DNA, the calcium in our teeth, the iron in our blood, the carbon in our apple pies were made in the interiors of collapsing stars. We are made of starstuff”~\cite{Sagan2002}. Therefore, for the time being we start the present bibliometric study from the Mendeleev Periodic Table as our initial platform which we shall connect later on with the astrochemistry as well as cosmochemistry.  

This investigation thus fundamentally identifies the main working field of research dynamics  and indicates the overall directional trend for the future research. As per the best of our knowledge, there is no such study having bibliometric analysis on the Mendeleev Periodic Table available in the literature as the stepping stone towards the astrochemical arena. Therefore, the following queries will be served as the pathways for the present study by using bibliometric analysis: (1) What is the current publication trend in the field of the Mendeleev Periodic Table? (2) Who are the most influential authors and relevant journals in this field of the Mendeleev Periodic Table? (3) Which are the most productive countries and institutions involved in the Mendeleev Periodic Table? (4) Who are the most prolific authors and what is the author collaborating trend in the Mendeleev Periodic Table? (5) Which are the most influential works in this field of the Mendeleev Periodic Table?

The outline of the present work is as follows: in the Section 2 we have provided background of the bibliometric analysis (Sub-Section 2.1) and the Mendeleev Periodic Table (Sub-Section 2.2). Section 3 is a precursor of the methodology we have adopted in the investigation. In Section 4 we have dealt with the results and findings on different features of the bibliometric analysis in the research field of the Mendeleev Periodic Table, e.g. Country production (Sub-section 4.1), Core sources by Bradford’s Law (Sub-section 4.2), Author’s local impact by h-Index (Sub-section 4.3), Author’s production (Sub-section 4.4), Annual scientific production (Sub-section 4.5), Most global cited documents (Sub-section 4.6), Co-occurrence network (Sub-section 4.7), Most relevant authors and countries (Sub-section 4.8). Section 5 is devoted for discussion and conclusion along with some future scope of the present study.

\section{Background history and overview}

\subsection{Brief history of bibliometric analysis}

The term {\it bibliometrics} was coined by Alan Pritchard in 1969~\cite{Pritchard1969}. However, in the late 19th century first appearance of bibliometrics studies occurred. The first instance of the term bibliometrics was found as its French equivalent hidden in a section titled ``Le Livre et la Mesure". The term bibliométric was first used by Paul Otlet in 1934~\cite{Otlet1934} and defined as ``the measurement of all aspects related to the publication and reading of books and documents"~\cite{Rousseau2014}. The anglicized version bibliometrics was first used by Alan Pritchard in a paper published in 1969, titled "Statistical Bibliography or Bibliometrics?"~\cite{Pritchard1969}. He defined the term as "the application of mathematics and statistical methods to books and other media of communication. The concept of bibliometrics "stresses the material aspect of the undertaking: counting books, articles, publications, citations"~\cite{Bellis2009}.

Bibliometric analysis appeared at the turn of the 19th and the 20th century~\cite{Hertzel2003,Godin2006,Danesh2020}. These developments predate the first occurrence of the concept of bibiometrics by several decades. Alternative label was commonly used: bibliography statistics became especially prevalent after 1920 and continued to remain in use until the end of the 1960s~\cite{Hertzel2003}. Early statistical studies of scientific metadata were motivated by the significant expansion of scientific output and the parallel development of indexing services of databases that made this information more accessible in the first place. Citation index was first applied to case law in the 1860s and their most famous example, Shepard's Citations (first published in 1873) will serve as a direct inspiration for the Science Citation Index one century later~\cite{Bellis2009}.

It is quantitative method that evaluates the inter-relationships and impacts of authors, institution, publications, and countries in a specific area. It is a kind of review method based on computer aided scientific data statistics. It uses mathematical and statistical tools to identify significant authors or studies and relationship between them~\cite{Ebrahim2017}.

\subsection{Brief overview of the Mendeleev Periodic Table}

~~~Dmitri Ivanovich Mendeleev, a Russian chemist was the most important donor to the early evolution of the periodic table. He is often referred as the Father of the Periodic Table. In 1869, after the spurning of Newland's Octave Law came into focus of the Mendeleev Periodic Table~\cite{Mendeleev2014}. In this, elements were aligned on the basis of their atomic mass and chemical properties. He arranged the elements in groups and periods. Vertical columns are referred as groups and horizontal row as periods. He found out that elements were arranged in increasing order of atomic mass and constant occurrence of elements with similar properties.

During Mendeleev's work, only 63 elements were investigated. After studying the properties of elements, he found that the properties of elements were related to atomic mass in a periodic way. He arranged the elements in such a pattern that elements with same properties come under same vertical column of the table. Mendeleev among chemical properties used formulae of hydrides and oxides as a basic criterion for classification.He took 63 cards and he wrote property of one element on each card. He makes a group of elements with similar properties and pinned it on the wall. He observed that elements were arranged in order of increasing atomic mass and elements with similar properties had periodic occurrence.

He articulated a periodic law based on his observations entitled ``The properties of elements are the periodic function of their atomic masses"~\cite{Mendeleev2014}. In his description of periodic table of 1871, he left gaps in which unknown elements would find their place in periodic table, believed by him. One of the demerits of this table is that increase in atomic mass was not regular while going from one element to other. Hence the number of elements yet to be discovered was not predictable. Two good review papers are available on the genesis of the Mendeleev Periodic Table in the following Refs.~\cite{Rawson1974,Kak2004}.

\section{Methodology}

~~~ It involves in two main procedures: performance analysis and science mapping analysis. The performance analysis investigates the activity indicators of publications, emphasizing the most significant contributions of research constituents to a given research field~\cite{Cristina2023}. On the other hand, science mapping analysis identifies and analyzes the scientific performance of articles, authors, institutions, countries and journals based on the number of citations, reveals the trends of the field studied through the analysis of keywords, identifies and clusters scientific gaps from most recent publications~\cite{Oliveira2019}.

It is a quantitative method that evaluates the inter-relationships and impacts of authors, institution, publications, and countries in a specific area. It is a kind of review method based on computer aided scientific review. It uses mathematical and statistical tools to identify significant authors or studies and relationship between them.

The study screened research paper keywords to identify important publications on Mendeleev periodic table. This study examined research papers on Mendeleev periodic table listed in the Scopus indexed journals because of its wider coverage of quality journals and its credibility of research information, rigorous indexing and large number of citations~\cite{Bergman2012}. We used ‘Mendeleev  periodic table’ as a keyword while screening the Scopus database. We comprehensively analysed all these papers by using bibliometric techniques and literature to examine the research papers on Mendeleev periodic table listed in the Scopus index. A bibliometric study is an analytical approach that uses the empirical and quantitative values to explain the distribution dynamics of research papers within a particular topic and a given period~\cite{Almind1997,Persson2009}.

 In this paper we set our limit of searching database to document type related to article and journal, language to English and subject area specified to chemistry, physics and astronomy and chemical engineering. We have selected only Q1 and Q2 based database in this paper.
 
 In the last two decades, bibliometric techniques have become a common method used in science and research.  Many areas of research use bibliometric methods for three reasons: expanding their area of operation, assessing the influence of a research group or determining the effect of a specific study ~\cite{Pilkington2009}. Therefore, they have become one of ‘the few truly interdisciplinary areas of research that can be extended to almost all fields of science that can be identified’~\cite{Glänzel2003}.

\section{Results and Findings}  

\subsection{Country production}
In Fig. 1, a graph has been drawn in which a plot between country production of articles over time period from 1959 to 2021 is discussed. Under this graph, we can infer from the data that USA has highest number of production of articles in the last year from 2015 to 2021. Unfortunately, Kazakhstan has no publication from 1959 to 2018. In the early years, i.e., from 1959 till 2003 there is no article publication by the scientists of Belgium, Brazil and China though later on they started publishing articles.
 
%%%%%%%%%%%%%%%%%%%%%%%%%%%%%%%%%%%%%%%%%%%%%%%%%%%%%%%%%%%%%%%%%%%%%%%%%%%%%%%%%%%%%%%%%%%
\begin{figure}
%\texttt{figure1.jpg}
    \includegraphics[scale=.10]{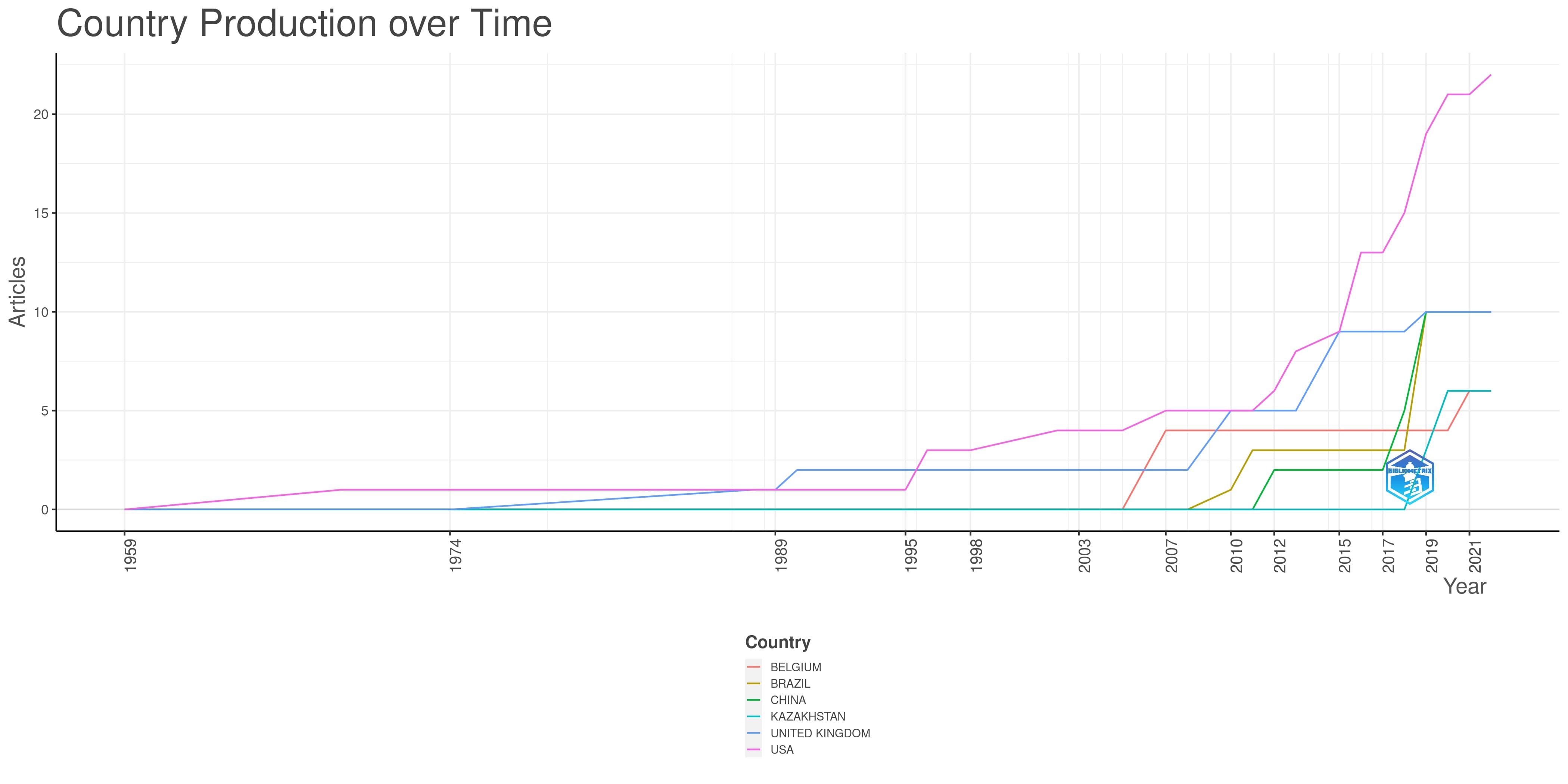}
    \caption{Country production over time}
    \label{fig:1}
\end{figure}
%%%%%%%%%%%%%%%%%%%%%%%%%%%%%%%%%%%%%%%%%%%%%%%%%%%%%%%%%%%%%%%%%%%%%%%%%%%%%%%%%%%%%%%%%%

 \subsection{Core sources by Bradford's Law}

 The Bradford Law can be defined as pattern that estimates the exponentially diminishing returns of searching for references in Science journals. This pattern was first described by Samuel C. Bradford in 1934~\cite{Bradford1934,Black2004,Yatsko2012}. This law is also known as Bradford's law of scattering or Bradford's distribution as it describes to show the articles of a particular subject that are scattered through the mass of periodicals as has been described: ``there are a few very productive periodicals, a larger number of more moderate producers, and a still larger number of constantly diminishing productivity”~\cite{Vickery1948,Hjorland2005}. Here a graph is plotted against number of articles and source rank of articles using Bradford's law (vide Fig. 2). It can be noted from the plot that the journal `Foundation of Chemistry' has the highest number of articles among others and `Crystallography Reports' has the least numbers.

  %%%%%%%%%%%%%%%%%%%%%%%%%%%%%%%%%%%%%%%%%%%%%%%%%%%%%%%%%%%%%%%%%%%%%%%%%%%%%%%%%%%%%%%%%%
 \begin{figure}
      \centering
      \includegraphics[scale=.10]{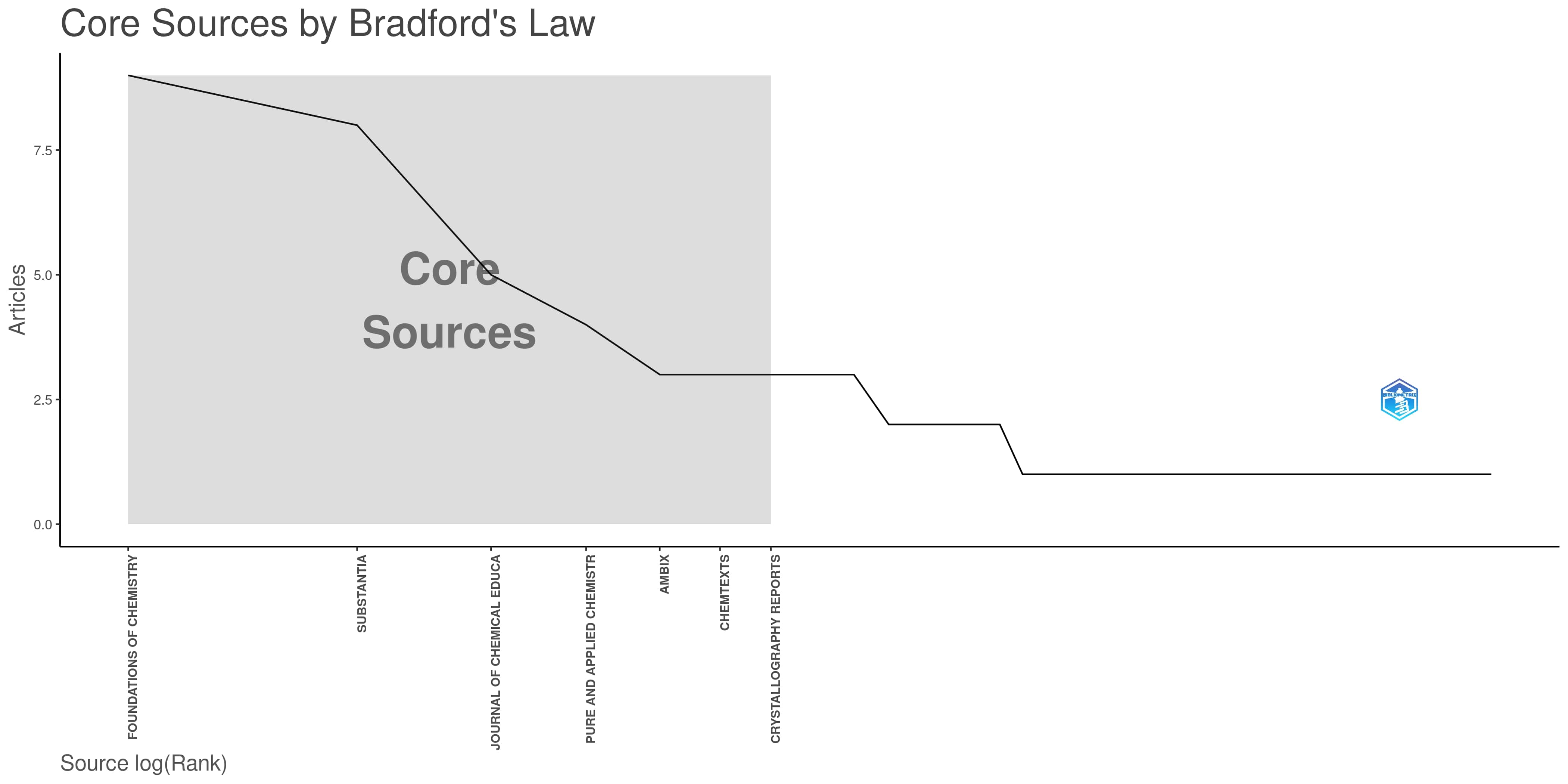}
      \caption{Core sources by Bradford's Law}
      \label{fig:2}
  \end{figure}   
 %%%%%%%%%%%%%%%%%%%%%%%%%%%%%%%%%%%%%%%%%%%%%%%%%%%%%%%%%%%%%%%%%%%%%%%%%%%%%%%%%%%%%%%%%%

\subsection{Author's local impact by $H$-index }

The $H$-index is an author-level matrix that measures both the productivity and citation impact of a publication~\cite{Hirsch2005}. Basically $H$-index can be attributed as a amazing tool for determining author's relative work impact by determining author's cited publication. From Fig. 3, it can be found out that Novaro has a $H$-index measure which is 2.0 based on the calculation technique. 
     
%%%%%%%%%%%%%%%%%%%%%%%%%%%%%%%%%%%%%%%%%%%%%%%%%%%%%%%%%%%%%%%%%%%%%%%%%%%%%%%%%%%%%%%%%%
    \begin{figure}
        \centering
        \includegraphics[scale=.10]{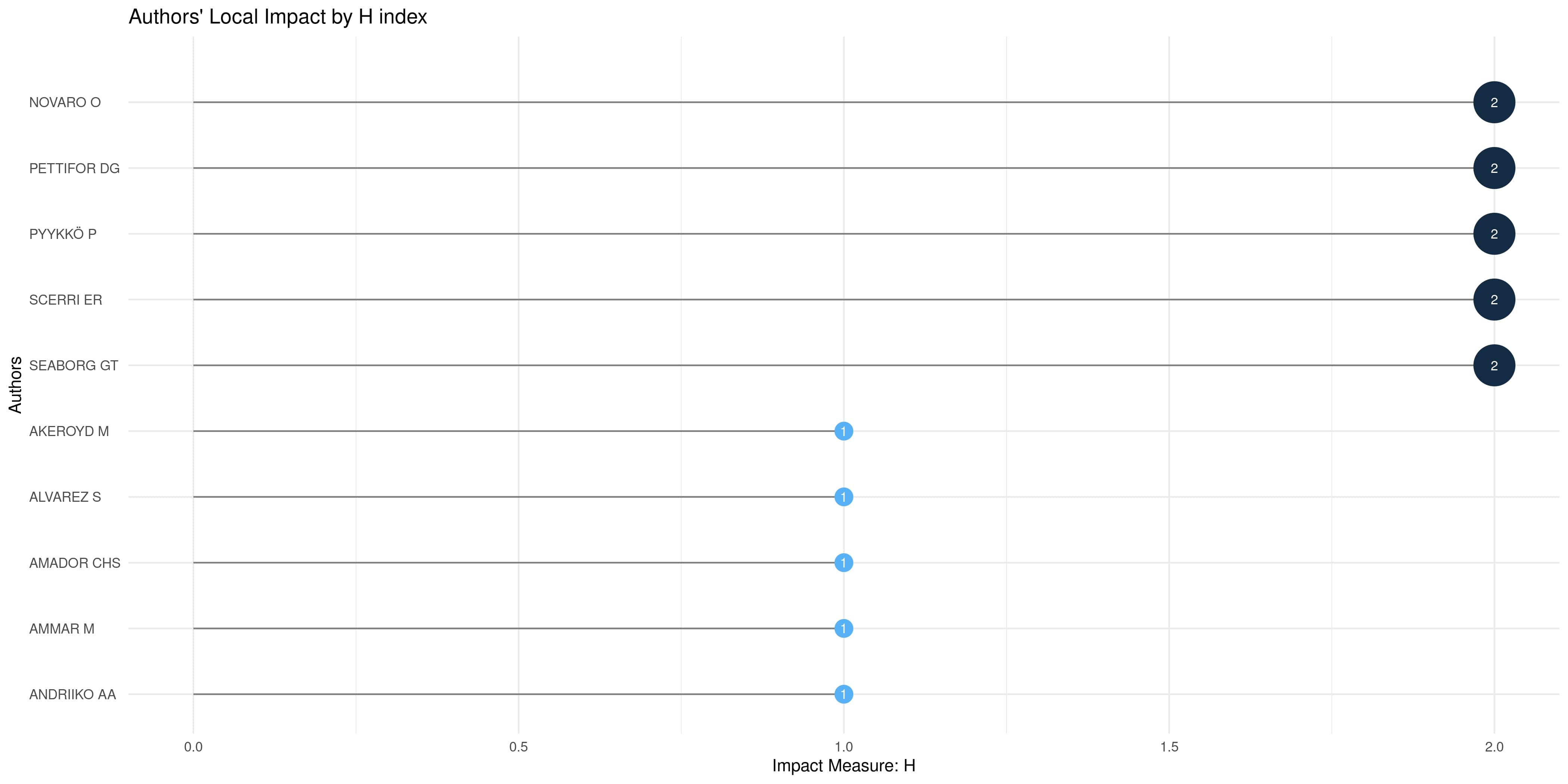}
        \caption{Author's local impact by H-Index}
        \label{fig:3}
    \end{figure}
%%%%%%%%%%%%%%%%%%%%%%%%%%%%%%%%%%%%%%%%%%%%%%%%%%%%%%%%%%%%%%%%%%%%%%%%%%%%%%%%%%%%%%%%%%

\subsection{Author's production}
   
It calculates and plots the author production in terms of number of publications over the time. There is a tool in bibliometrix  package for $R$ which plots number of publication of authors over period of time. This tool takes input of bibliographic data and returns a list containing two data frames. 
     
The first data frame consists of author's name and their publication count over time, while the second frame consists of publication count of each author.This tool can be used to know the productivity of author's in field of interest. Figure 3 indicates that the author O. Novaro has the highest production over time period of 1989 to 2008.  
     
%%%%%%%%%%%%%%%%%%%%%%%%%%%%%%%%%%%%%%%%%%%%%%%%%%%%%%%%%%%%%%%%%%%%%%%%%%%%%%%%%%%%%%%%%%
     \begin{figure}
         \centering
         \includegraphics[scale=.08]{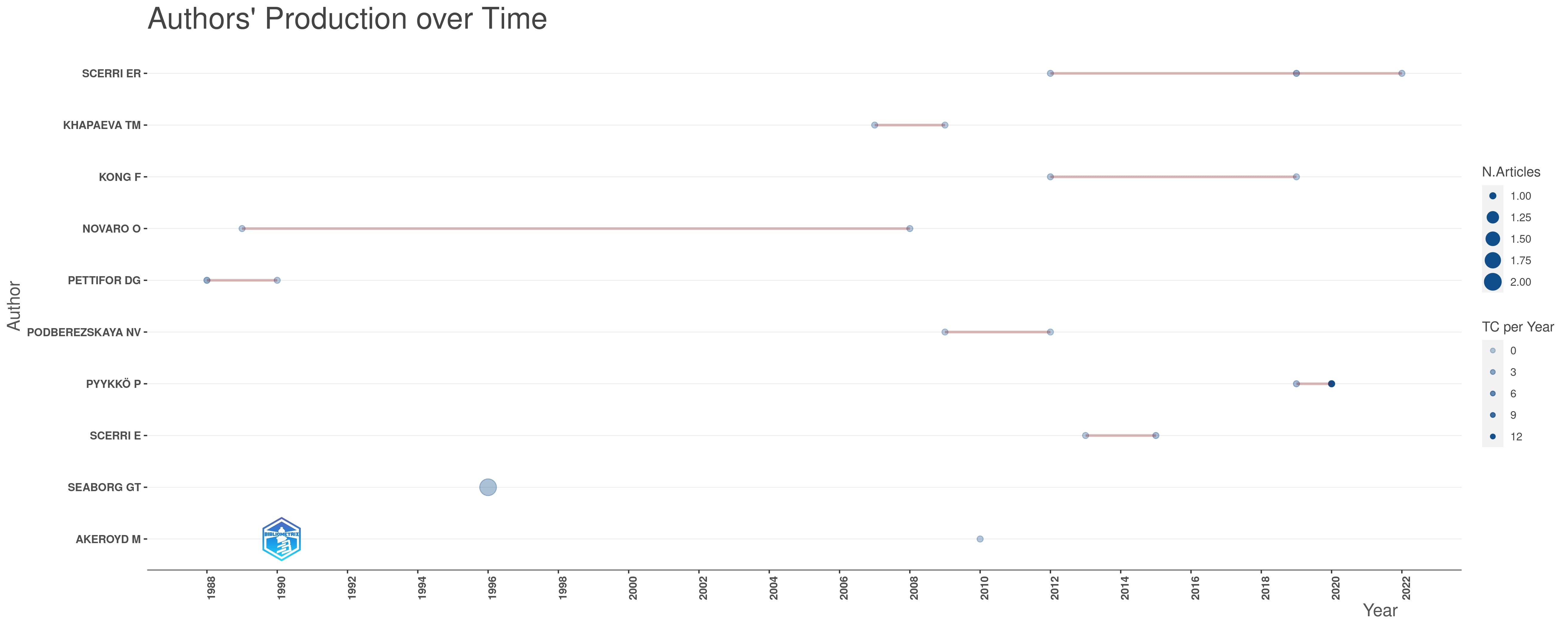}
         \caption{Author's production over the time}
         \label{fig:4}
     \end{figure}
%%%%%%%%%%%%%%%%%%%%%%%%%%%%%%%%%%%%%%%%%%%%%%%%%%%%%%%%%%%%%%%%%%%%%%%%%%%%%%%%%%%%%%%%%%

\subsection{Annual scientific production}

A plot is drafted between number of articles and year time period ranging from 1959 to 2021. The peak is seen at year 2019 which gives the idea of the highest number of articles in that particular year. A sudden rise in articles from 2017 to 2019 is observed. This plot of annual scientific production depicts the era production of articles on Mendeleev Periodic Table. In the early year's scientific production is very less which rises after 2017.
     
%%%%%%%%%%%%%%%%%%%%%%%%%%%%%%%%%%%%%%%%%%%%%%%%%%%%%%%%%%%%%%%%%%%%%%%%%%%%%%%%%%%%%%%%%%  
     \begin{figure}
         \centering
         \includegraphics[scale=.10]{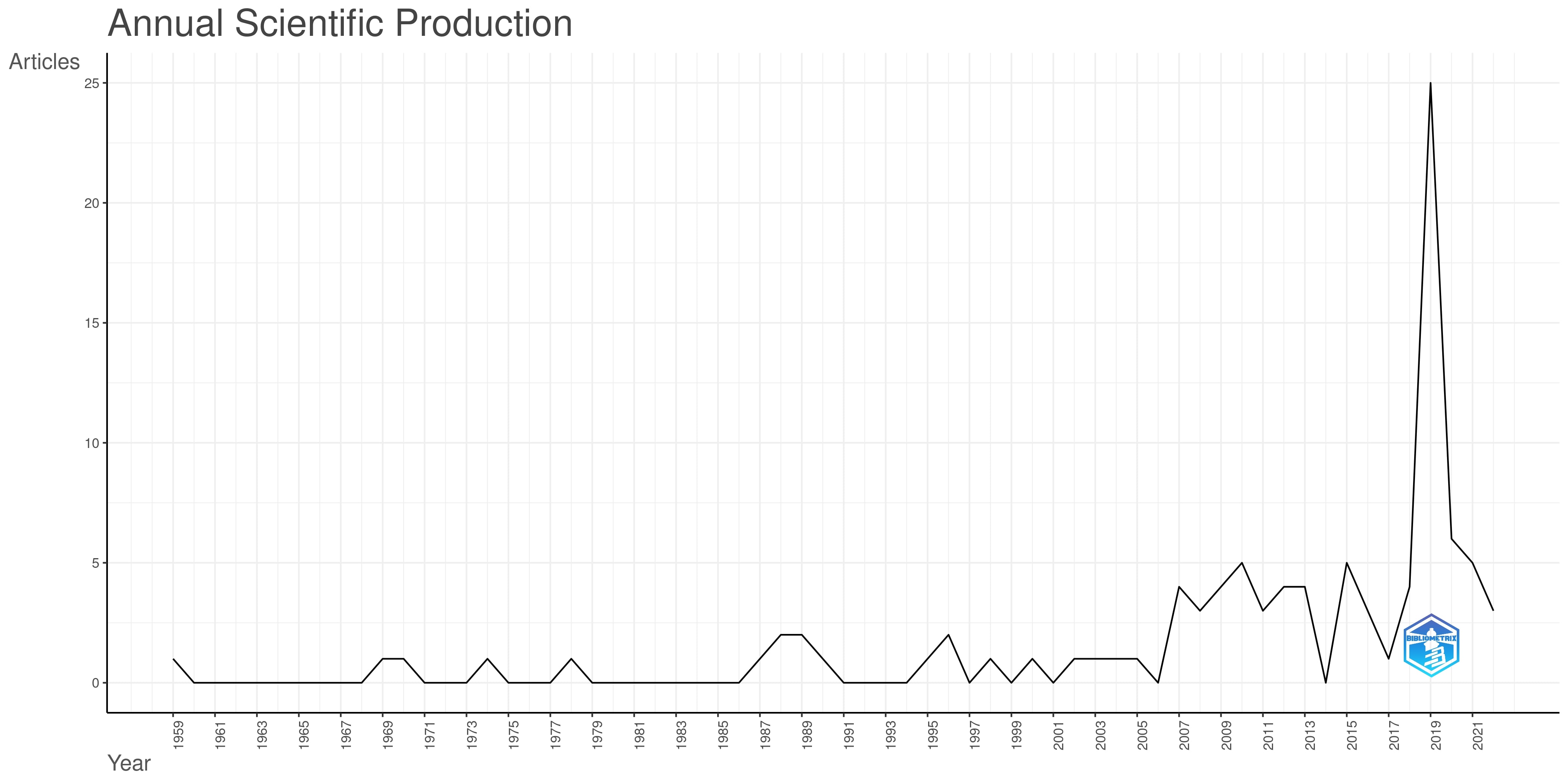}
         \caption{Annual scientific production}
         \label{fig:5}
     \end{figure}
%%%%%%%%%%%%%%%%%%%%%%%%%%%%%%%%%%%%%%%%%%%%%%%%%%%%%%%%%%%%%%%%%%%%%%%%%%%%%%%%%%%%%%%%%%

\subsection{Most global cited documents }
     
According to a bibliometrix analysis based on the number of publications, keywords and citations in Mendeleev's Periodic Table, the most globally cited document with 156 citations was the paper entitled ``Electron affinities of the heavy elements" by R.J. Zollweg in 1969~\cite{Zollweg1969}. On the other hand, the least cited document is ``Superheavy elements in D I Mendeleev's periodic table" by Y.T. Oganessian and S.N. Dmitriev in 2009 with only 22 citations~\cite{Oganessian2009}. This feature is very obvious as over time the paper by Zollweg received much citation in comparison to the work by Oganessian and Dmitriev. 
     
 %%%%%%%%%%%%%%%%%%%%%%%%%%%%%%%%%%%%%%%%%%%%%%%%%%%%%%%%%%%%%%%%%%%%%%%%%%%%%%%%%%%%%%%%%%
     \begin{figure}
         \centering
         \includegraphics[scale=.10]{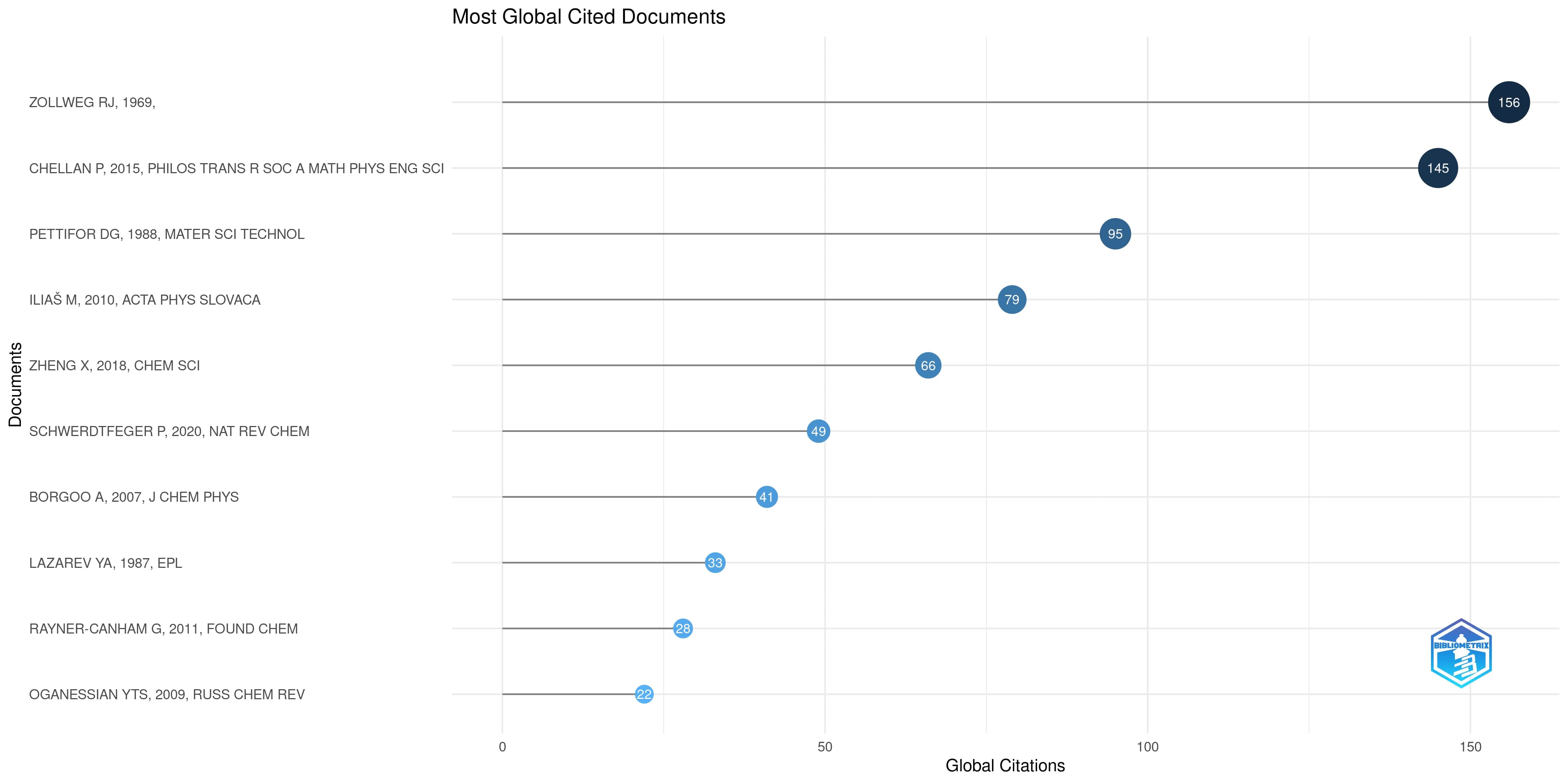}
         \caption{Most global cited documents over the time}
         \label{fig:6}
     \end{figure}
%%%%%%%%%%%%%%%%%%%%%%%%%%%%%%%%%%%%%%%%%%%%%%%%%%%%%%%%%%%%%%%%%%%%%%%%%%%%%%%%%%%%%%%%%%

\subsection{Co-occurrence network}
  
Co-occurrence network sometime referred as semantic network. It can be attributed as an useful tool to analyse the big data and large text by identifying the main themes and topics, or even mapping an entire research field. The method of constructing co-occurrence networks includes identifying keywords in the text, calculating the frequencies of co-occurrence and analysing the networks. Map represents the connection of various keywords to  each other with lines. These lines signify the co-occurrence of keywords with each other in different papers in the considered data set collected from Scopus database. It is to be distinctively noted here that `Periodic Table' is the main keyword which in turn related to atoms, molecules, chemical elements, bonds, electrons and so on which are co-related to each other.
    
%%%%%%%%%%%%%%%%%%%%%%%%%%%%%%%%%%%%%%%%%%%%%%%%%%%%%%%%%%%%%%%%%%%%%%%%%%%%%%%%%%%%%%%%%% 
     \begin{figure}
         \centering
         \includegraphics[scale=.30]{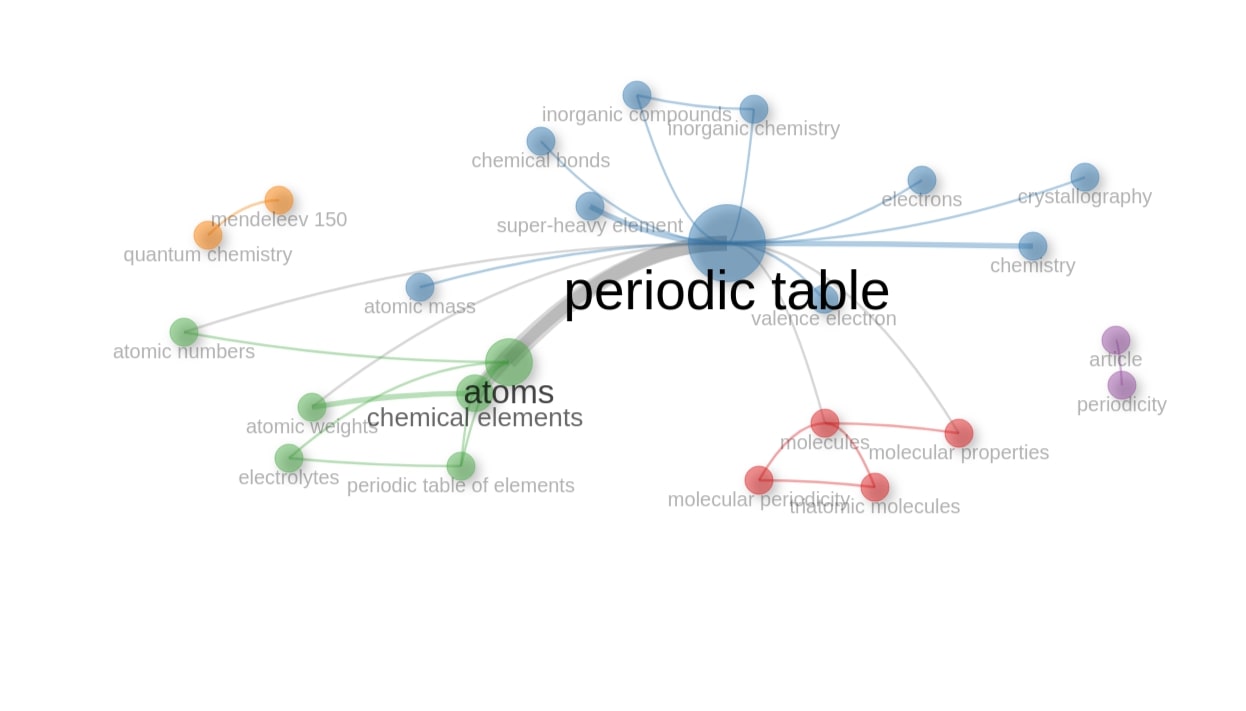}
         \caption{Co-occurrence Network over the time}
         \label{fig:7}
     \end{figure}
%%%%%%%%%%%%%%%%%%%%%%%%%%%%%%%%%%%%%%%%%%%%%%%%%%%%%%%%%%%%%%%%%%%%%%%%%%%%%%%%%%%%%%%%%%     

\subsection{Most relevant authors and countries}
    
%%%%%%%%%%%%%%%%%%%%%%%%%%%%%%%%%%%%%%%%%%%%%%%%%%%%%%%%%%%%%%%%%%%%%%%%%%%%%%%%%%%%%%%%%% 
     \begin{figure}
         \centering
         \includegraphics[scale=.10]{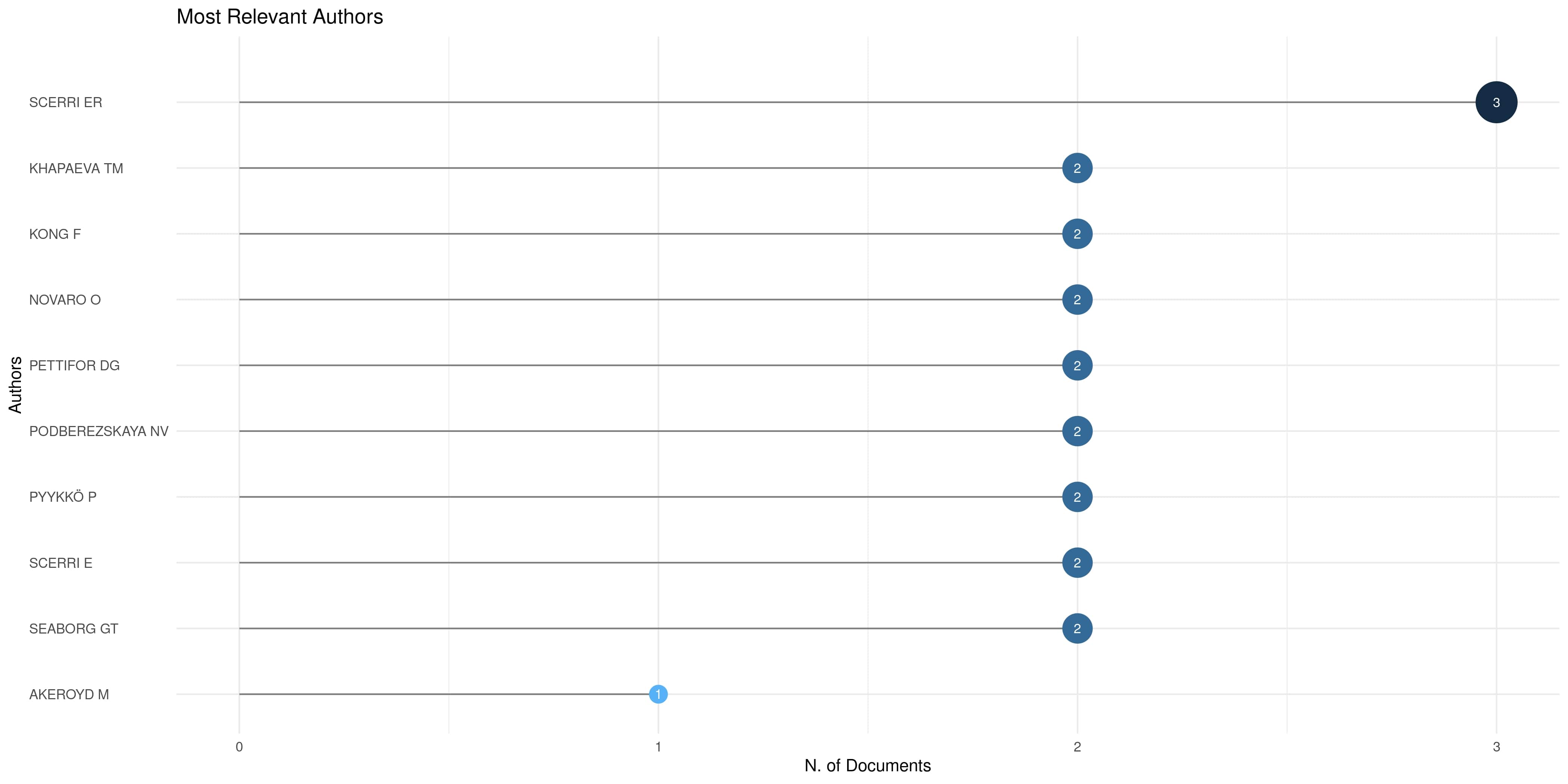}
         \caption{Most relevant authors over the time}
         \label{fig:8}
     \end{figure}
     
     \begin{figure}
         \centering
         \includegraphics[scale=.10]{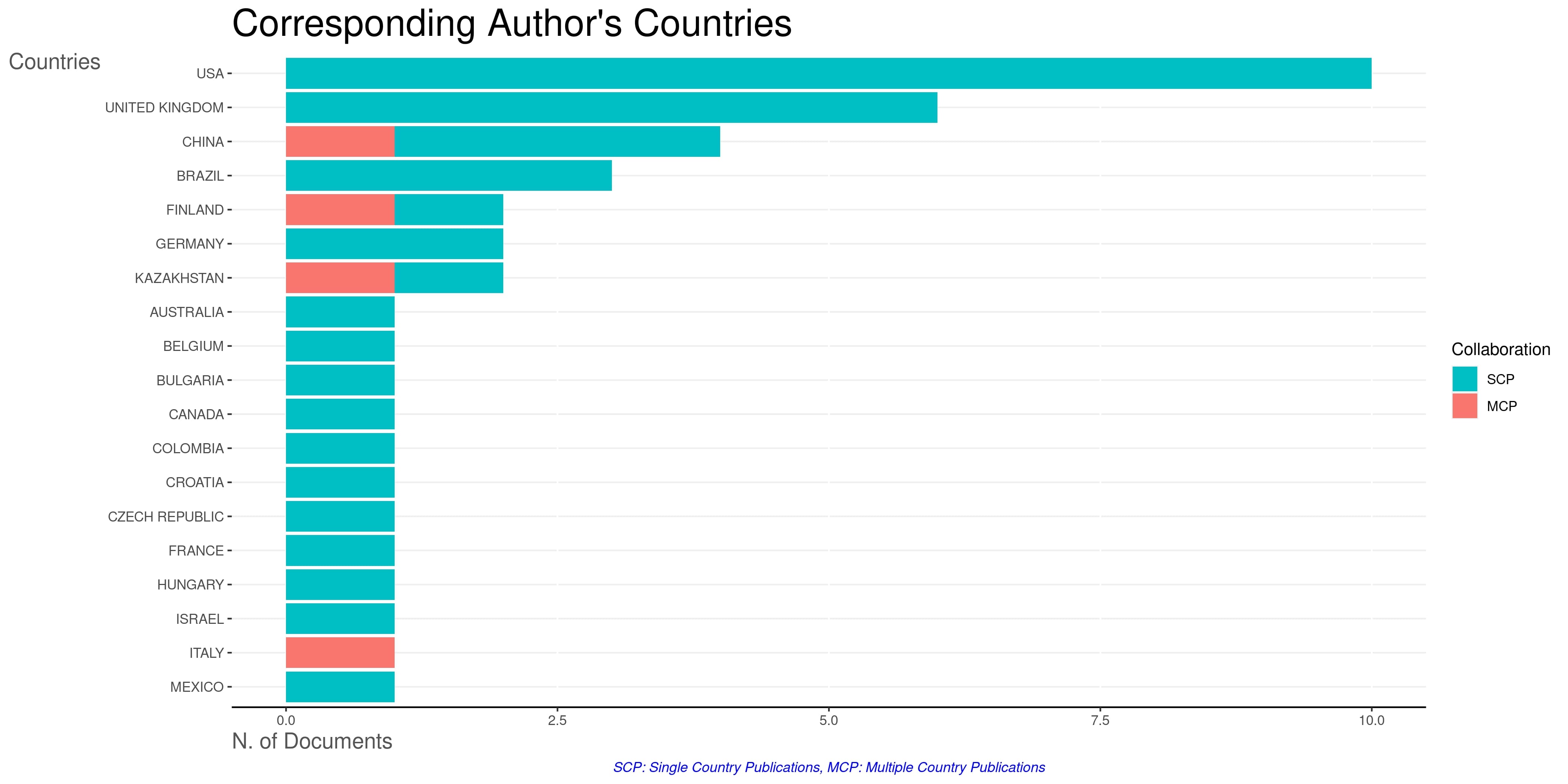}
         \caption{Most relevant countries over the time}
         \label{fig:9}
     \end{figure}  
%%%%%%%%%%%%%%%%%%%%%%%%%%%%%%%%%%%%%%%%%%%%%%%%%%%%%%%%%%%%%%%%%%%%%%%%%%%%%%%%%%%%%%%%%%

In the Fig. 8, authors are placed in an order according to decrease in the number of documents. E.R. Scerri is the most relevant author with three documents based on the data set collected from Scopus database whereas in the bottom of the figure there is an entry with the name M. Akeroyd. 

On the other hand, the database collected from Scopus and analysed with the bibliometrix tool provides the most corresponding author's countries on the basis of their publication on Mendeleev's periodic table (Fig. 9). Countries can publish their data in two ways: single country publication and in a collaboration with different countries as multiple country publication. USA has the highest number of publications as single country publication among others countries. Italy has only multiple country publication collaboration. China, Brazil and Kazakhstan all three countries have both the of publications, i.e., single country publication and multiple country publication.

\section{Discussion and Conclusion}

This study represents the bibliometric analysis of various studies published on Mendeleev Periodic Table. We have selected Q1 and Q2 from the Scopus database. This study focused on the year wise trends of publication of Mendeleev Periodic Table. It identifies the most global cited documents, most relevant authors and most relevant author's countries and authors productivity over time that are prominently involved in its research. In addition to, this study presents co-occurrence network analysis, annual scientific production and author's impact by $h$-index. \\

Some of the salient features of the present investigations are as follows:

(i) These analyses give a direction to researchers venturing into the area of Mendeleev's Periodic Table by providing information on journals, authors, collaborators and countries that are prominent in this domain as well as co-occurrence network used in the research. 

(ii) From the data we analysed we can infer that USA is the country that has the highest publication over time. One can observe from the relevant plot (vide Fig. 3) that Novaro is the highest productive author of time period 1989 to 2008. Also, it is evident that the year 2019 has the highest annual scientific publication in Mendeleev Periodic Table. 

(iii) The most globally cited document in the field of Mendeleev 's Periodic Table is ``Electron affinity of heavy elements" with 156 citations published in the ``Journal of Chemical Physics". E.R. Scerri is the most relevant author of the time with the highest number of documents on periodic table. USA can be awarded as the most relevant country in term of publication.  Results of annual scientific production and the most cited document have collected increasing interest in the field. It is important for the future researchers to explore the ideas that are needed to identify the research gap. The analysis of relevant author's and countries indicates global interest in this domain. 

(iv) Research can be attributed more powerful and impacting effect when it gives a global result. Such studies enable research students as well as scientists to collaborate with experts who are working globally. These results are valuable for finding research journals, articles and other issues present in the domain of periodic table in that time span which in turn will be beneficial for the coming generation to find out possible outcomes in the field of research. 

In connection to the scope for future study we would like to state here that in this paper, we only included the data which is based on Scopus database of Mendeleev's Periodic Table containing Q1 and Q2 factor so the paper which are not indexed in Scopus are not included in research. Therefore, (i) the future researcher can collect more amount of data from other sources such as Web of Science and Google Scholar, (ii) more comprehensive research can be carried out by focusing on different network diagrams that are not studied out here, and (iii) a diversified study can be done by extending the time span of database selection where the data set can be studied comprehensively by including different network. Furthermore, one can study the astrochemistry problems based on the present structure where micro-chemical elements can be connected to the macro-objects under the periphery of cosmological realm. This aspect may be considered in one of our future projects on the astrochemistry research which we have started here via the Mendeleev Periodic Table. 

 We can summarise that bibliometric analysis is an influential tool to explore publishing trends and their relationship between communicated work. Furthermore, it helps in recognizing the  most influential researches as well as researcher in the field as evident in the present study of the Mendeleev Periodic Table.

\section*{Declaration of competing interest}
The authors declare that they have no known competing financial interests or personal relationships that could have appeared
to influence the work reported in this paper

\bmhead{Acknowledgements}
SR gratefully acknowledges support from the Inter-University Centre for Astronomy and Astrophysics (IUCAA), Pune, India under its Visiting Research Associateship Programme and for the facilities under ICARD, Pune at CCASS, GLA University, Mathura. \\

\section*{Declarations}

\begin{itemize}
\item Funding : No funds, grants, or other support was received.
\item Conflict of interest/Competing interests (check journal-specific guidelines for which heading to use) : The authors have no competing interests to declare that are relevant to the content of this article.
\item Ethics approval : Not applicable
\item Consent to participate : Not applicable
\item Consent for publication : Yes
\item Availability of data and materials : A comprehensive data archive is available as open source in the Internet.
\item Code availability : Not applicable
\item Authors' contributions : All the authors contributed equally to this work.
\end{itemize}

\end{document}